\documentclass[a4paper, amsfonts, amssymb, amsmath, reprint, prl, aps,
         showkeys, nofootinbcib, superscriptaddress, twoside, twocolumn]{revtex4-2}

\usepackage[makeroom]{cancel}
\usepackage{dcolumn}
\usepackage{makecell}
\usepackage[utf8]{inputenc}
\usepackage[T1]{fontenc}
\usepackage{mathptmx}
\usepackage{commath}
\usepackage{graphicx}
\usepackage{multirow}
\usepackage{amsmath}

\usepackage{soul}
\usepackage{verbatim}
\usepackage{bm}
\usepackage{outlines}
\usepackage{siunitx}
\usepackage[space]{grffile}
\usepackage{upgreek}
\usepackage[dvipsnames]{xcolor}
\usepackage{xr}
\usepackage{color}
\usepackage{tikz}
\usepackage{mathtools}
\usepackage{amssymb,placeins}
\usepackage[globalcitecopy]{bibunits}
\usetikzlibrary{quantikz}

\newcommand\myshade{85}
\newcommand{\blockmatrix}[1]{\begin{bmatrix} #1 \end{bmatrix}}
\colorlet{mylinkcolor}{violet}
\colorlet{mycitecolor}{YellowOrange}
\colorlet{myurlcolor}{Aquamarine}
\usepackage{hyperref}
\hypersetup{
  linkcolor  = mylinkcolor!\myshade!black,
  citecolor  = mycitecolor!\myshade!black,
  urlcolor   = myurlcolor!\myshade!black,
  colorlinks = true,
}
\usepackage[capitalize]{cleveref}
\usetikzlibrary{calc, positioning}

\newcommand{\QuTech}{\affiliation{QuTech and Kavli Institute of Nanoscience, Delft University of Technology, P.O. Box 5046, 2600 GA Delft, The Netherlands}}
\newcommand{\OQS}{\affiliation{Orange Quantum Systems, Elektronicaweg 2, 2628 XG Delft, The Netherlands}}
\newcommand{\QW}{\affiliation{Quantware B.V., Elektronicaweg 10, 2628 XG Delft, The Netherlands}}

\newcommand{\nametitle}{Optimizing the frequency positioning of tunable couplers in a circuit QED processor to mitigate spectator effects on quantum operations}

\newcommand{\qa}{\textrm{Q}_{\textrm{1}}}
\newcommand{\qb}{\textrm{Q}_{\textrm{2}}}
\newcommand{\qc}{\textrm{Q}_{\textrm{0}}}
\newcommand{\qd}{\textrm{Q}_{\textrm{3}}}
\newcommand{\qe}{\textrm{Q}_{\textrm{4}}}
\newcommand{\qi}{\textrm{Q}_{i}}
\newcommand{\qj}{\textrm{Q}_{j}}
\newcommand{\cca}{\textrm{C}_{\textrm{01}}}
\newcommand{\ccb}{\textrm{C}_{\textrm{02}}}
\newcommand{\ccd}{\textrm{C}_{\textrm{03}}}
\newcommand{\cce}{\textrm{C}_{\textrm{04}}}
\newcommand{\cij}{\textrm{C}_{ij}}
\newcommand{\cci}{\textrm{C}_{0i}}
\newcommand{\jone}{J_{\textrm{1}}}
\newcommand{\jtwo}{J_{\textrm{2}}}
\newcommand{\reszz}{\xi_{ZZ}}
\newcommand{\pihalf}{\mathrm{Y}_{\mathrm{\pi}/2}}
\newcommand{\tauopt}{\tau_{\mathrm{opt}}}
\newcommand{\gqc}{g_{\mathrm{qc}}}
\newcommand{\gqq}{g_{\mathrm{qq}}}

\newcommand{\epsro}{\epsilon_{\mathrm{RO}}}
\newcommand{\epscz}{\epsilon_{\mathrm{CZ}}}
\newcommand{\lone}{L_{1}}
\newcommand{\us}{\upmu \mathrm{s}}
\newcommand{\ns}{\mathrm{ns}}

\newcommand{\MHz}{\mathrm{MHz}}
\newcommand{\kHz}{\mathrm{kHz}}

\newcommand{\mK}{\mathrm{mK}}
\newcommand{\K}{\mathrm{K}}
\newcommand{\minutes}{\mathrm{min}}
\newcommand{\degrees}{\circ}

\begin{document}

\title{\nametitle}
\author{S.~Vallés-Sanclemente}\QuTech
\author{T.~H.~F.~Vroomans}\QuTech
\author{T.~R.~van~Abswoude}\OQS
\author{F.~Brulleman}\QW
\author{T.~Stavenga}\QW
\author{S.~L.~M.~van~der~Meer}
\author{Y.~Xin}\QuTech
\author{A.~Lawrence}\OQS
\author{V.~Singh}\thanks{Present address: A*STAR Quantum Innovation Centre (Q.InC), Institute of Materials Research and Engineering (IMRE), Agency for Science, Technology and Research (A*STAR), 2 Fusionopolis Way, 08-03, Singapore, Republic of Singapore}\QuTech
\author{M.~A.~Rol}\OQS
\author{L.~DiCarlo}\QuTech

\date{\today}

\begin{abstract}
We experimentally optimize the frequency of flux-tunable couplers in a superconducting quantum processor to minimize the impact of spectator transmons during quantum operations 
(single-qubit gates, two-qubit gates and readout) on other transmons. We adapt a popular transmon-like tunable-coupling element, achieving high-fidelity, low-leakage controlled-$Z$ gates with unipolar, fast-adiabatic pulsing only on the coupler. We demonstrate the ability of the tunable coupler to null residual $ZZ$ coupling as well as exchange couplings in the one- and two-excitation manifolds. However, the nulling of these coherent interactions is not simultaneous, prompting the exploration of tradeoffs. We present experiments pinpointing spectator effects on specific quantum operations. We also study the combined effect on the three types of operations using repeated quantum parity measurements.
\end{abstract}
\maketitle

\begin{bibunit}[apsrev4-2]

Residual quantum interactions between qubits are a key source of crosstalk in multi-qubit processors. In particular, spectator (i.e., inactive) qubits during single-qubit gates, two-qubit gates and readout of other qubits can introduce state-dependent errors. In legacy circuit QED processors employing transmons coupled via fixed-frequency resonators, a dominant residual interaction is $ZZ$ coupling, which adds a Hamiltonian term
\[
\delta H = \hbar\reszz^{ij}\ket{1_{i}1_{j}}\bra{1_{i}1_{j}}, \\
\]
making the qubit transition frequency of transmon $\qi$ dependent on the state of $\qj$, and viceversa. (Here, $\hbar$ is the reduced Planck constant and numbers indicate energy excitation level.) It is not uncommon for the coupling strength $|\reszz|/(2\pi)$ to reach $\MHz$-level for some pairs~\cite{Marques22}, due to limited frequency targeting. Post-fabrication frequency trimming methods have allowed taming $|\reszz|/(2\pi)$ to the $\sim100~\kHz$ range in many cases~\cite{Zhang22}.

Exchange interactions in the one- and two-excitation manifolds are strong, adding Hamiltonian terms
\begin{align*}
&\hbar\jone^{ij}\ket{0_{i}1_{j}}\bra{1_{i}0_{j}} + \mathrm{h.c.}, \\
&\hbar\jtwo^{ij}\ket{1_{i}1_{j}}\bra{0_{i}2_{j}} + \mathrm{h.c.},
\end{align*}
with coupling strengths $J_{1,2}/(2\pi)$ typically at the $10-\MHz$ level. Only large detuning $|\Delta| \gg J$ between the coupled levels limits their impact.

In recent years, significant effort has focused on the development of flux-tunable couplers to mitigate these interactions~\cite{Hime06, Bertet06, Niskanen07, VanDerPloeg07, Harris07, Allman10, Bialczak11, Srinivasan11, Yin13, Peropadre13, Hoi13, Pierre14, Chen14b, McKay16, Weber17, Lu17, Kounalakis18}. A popular approach uses flux-tunable transmon-like elements~\cite{Koch07, Yan18, Rosenblum18, Gao19, Mundada19, Arute19, Collodo20, Foxen20, Xu20, Li20b, Han20, Google21, Sung21b, Stehlik21, Sete21, Ye21, Zhao22c, Moskalenko22, Goto22, Marxer23, Kubo23, Campbell23, Google23, Ding23, Krivzan24, Kubo24, Google25, Liu25, Zhang25, Li25}, with variants including floating~\cite{Sete21, Marxer23}, grounded~\cite{Sung21b} and chiplet-linking configurations~\cite{Campbell23}. A characteristic of this approach is that $\reszz$, $\jone$ and $\jtwo$ can all be nulled, provided the coupled transmons lie in the straddling regime where the qubit $\left(\ket{0}-\ket{1}\right)$ transition of one transmon is flanked by the qubit and leakage $\left(\ket{1}-\ket{2}\right)$ transitions of the other. Crucially though, $\reszz$, $\jone$ and $\jtwo$ cannot be simultaneously nulled at the same frequency positioning of the coupler. This fact motivates the question: what tunable coupler positioning is best to mitigate the impact of spectator transmons on the fidelity of quantum operations (readout, single- and two-qubit gates) performed on other transmons~\cite{Sung21b}?

In this Letter, we experimentally address this question using a 5-transmon prototype of our tunable-coupler-based processor architecture under development. After mapping out the dependence of $\reszz$, $\jone$ and $\jtwo$ on coupler frequency, we first find that spectator effects on single-qubit gates are strongly mitigated over a wide range of coupler positioning, spanning both $\reszz=0$ and $\jone=0$, but not $\jtwo=0$. However, for readout, nulling $\jone$ or $\jtwo$ is advantageous: it avoids the excitation exchange activated by measurement-induced spectral overlap between transitions of coupled transmons. To also consider the impact of spectators on two-qubit gates, we first consider several options for realizing controlled-$Z$ (CZ) gates, converging on  unipolar, fast-adiabatic flux pulsing only on the tunable coupler, achieving CZ error $\epscz<1\%$ with leakage $\lone\approx 0.05\%$. Our final experiment studies the combined effect of spectators on readout, single- and two-qubit gates in a minimal setting relevant for quantum error correction (QEC): repeated weight-2 parity checks, which yield a baseline defect rate of $\sim12\%$. By repeatedly toggling a spectator, we observe that the defect rate is lowest with the coupler bias in a region that includes $\reszz=0$, while it degrades at $\jone=0$ and $\jtwo=0$. For ultimate performance, it may be worthwhile to dynamically switch coupler positioning for readout and gate operations, taking advantage of the fast flux tunability already in place.

\begin{figure}
\centering
\includegraphics{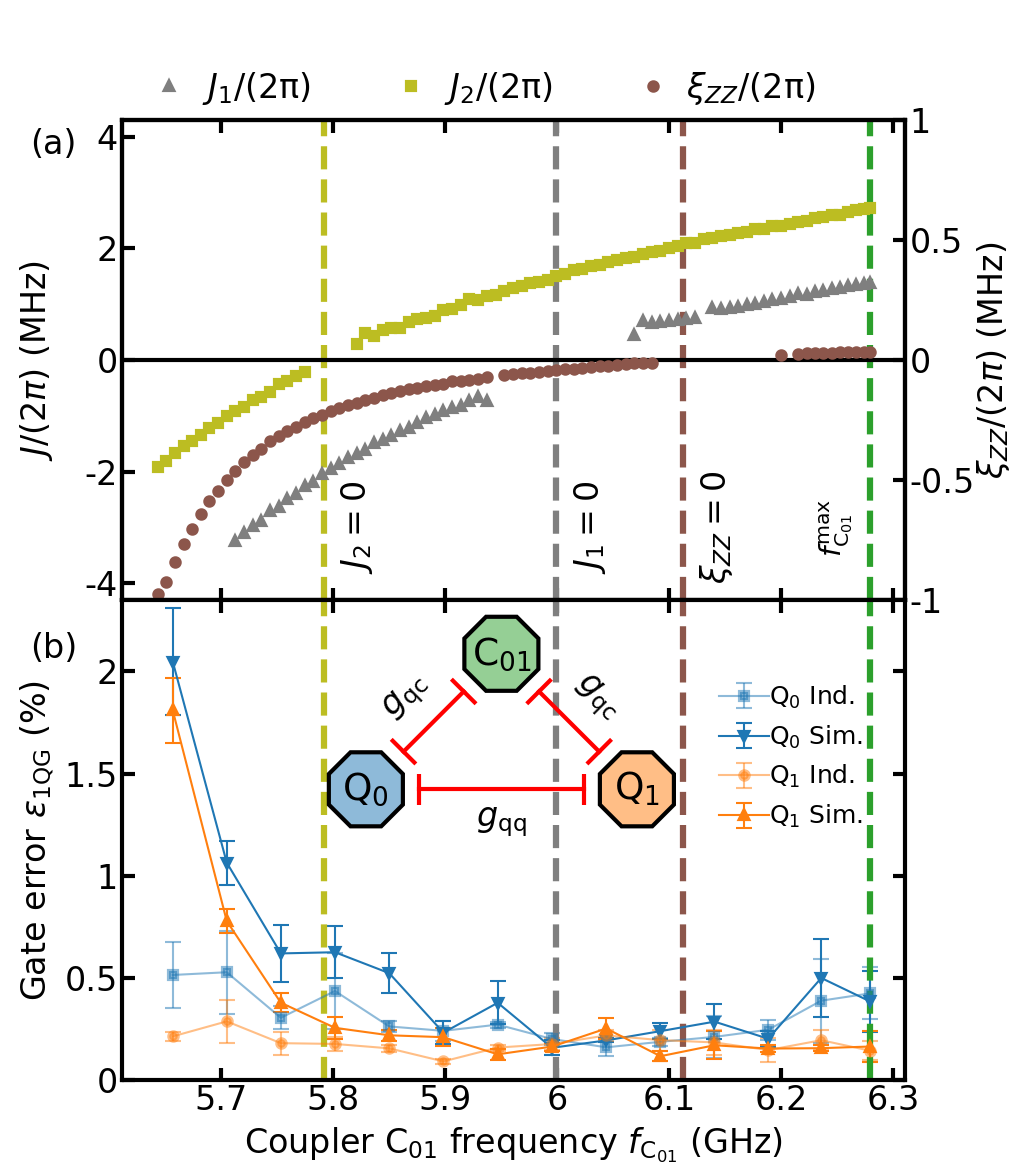}
\caption{
Single-qubit gate performance as a function of tunable coupler frequency. (a) $\jone$, $\jtwo$ and $\reszz$ between $\qa$ and $\qc$ as a function of coupler $\cca$ frequency. Vertical grey, golden and brown lines indicate the $\cca$ frequencies at which $\jone$, $\jtwo$ and $\reszz$ are nulled, respectively. (b) Single-qubit gate errors on $\qc$ and $\qa$, measured using simultaneous and individual randomized benchmarking, as a function of $\cca$ frequency. Error bars are the standard deviation of the mean computed from 10 repetitions. Inset: Schematic of the sub-system formed by $\qa$, $\qc$ and $\cca$.
\label{fig:SingleQubitGates}}
\end{figure}

\begin{figure}
\centering
\includegraphics{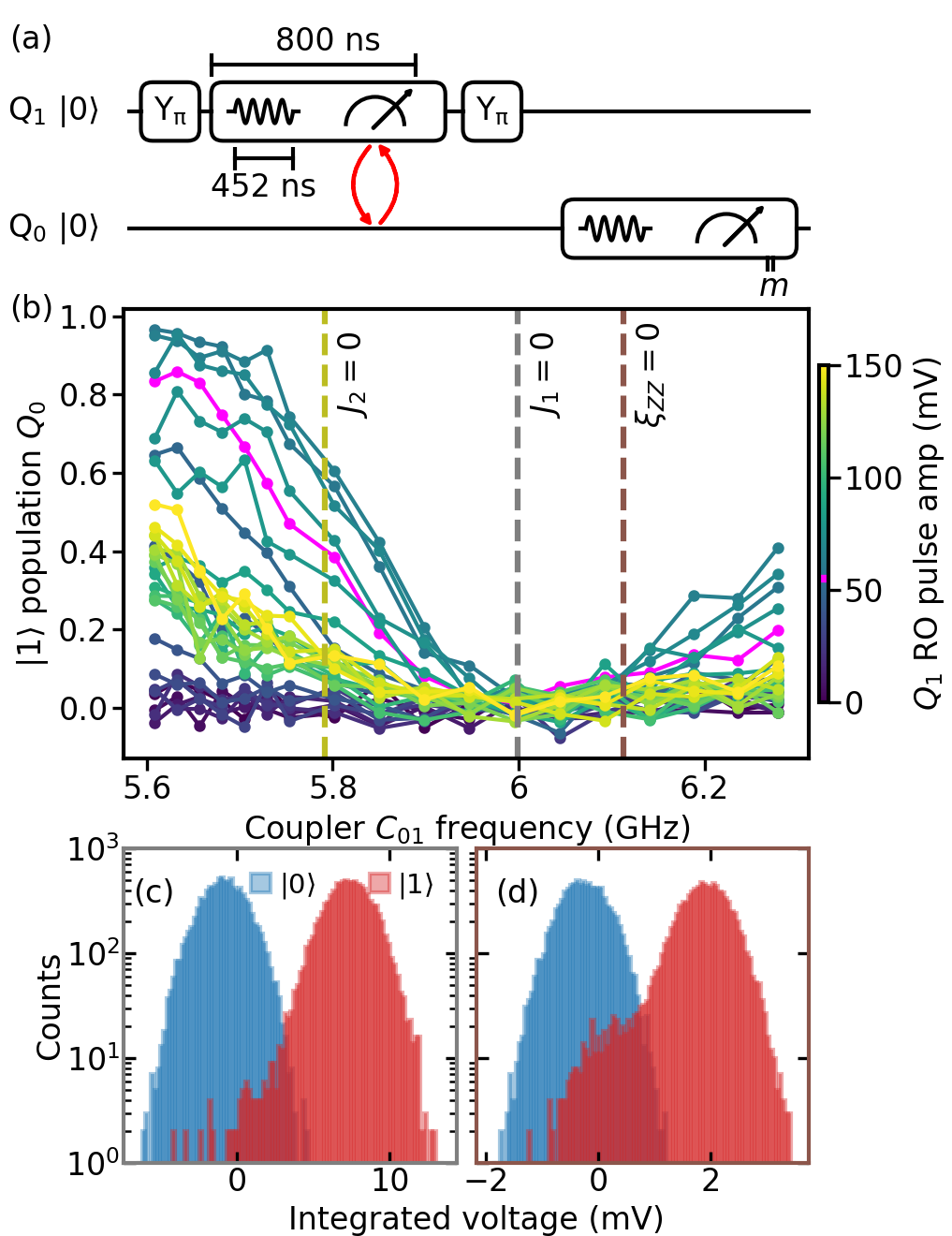}
\caption{
Measurement-induced population exchange. (a) Experiment used to study  population exchange from $\qc$ to $\qa$ during readout of $\qc$. (b) Population of $\qa$ as a function of $\cca$ frequency and amplitude of the $\qc$ readout pulse. The pink trace shows the data acquired with the chosen amplitude~\cite{SOM_tc_spect}. (c) Single-shot readout (SSRO) characterization for $\qc$ at $\jone=0$, yielding $\epsro=0.8\%$. (d) Similar characterization at $\reszz=0$ and using identical readout parameters, yielding $\epsro=2.5\%$.
\label{fig:J1ROPExch}}
\end{figure}

Our device consists of five flux-tunable transmon qubits in a starfish configuration, with central transmon $\qc$ connected to four corner transmons ($\qa$-$\qe$) via dedicated, also flux-tunable couplers $\cci$ (Please see~\cite{SOM_tc_spect} for device parameters and performance characteristics). Numerical simulations show that the targeted coupling strengths, $\gqq/(2\pi)=6~\MHz$ and $\gqc/(2\pi)=70~\MHz$~\cite{SOM_tc_spect}, can null $\reszz$ with the transmons in the so-called straddling regime and the coupler positioned $\sim700~\MHz$ above the qubit transition frequencies. In this device, we find all coupled transmon pairs in the straddling regime with all biased at their simultaneous sweetspot, making this the clear choice of bias point for transmons as it maximizes qubit coherence.

We start our study by mapping the dependence of $\reszz$, $\jone$ and $\jtwo$ on coupler bias frequency. Figure~\ref{fig:SingleQubitGates}(a) shows example results for pair $(\qa,\qc)$ (see~\cite{SOM_tc_spect} for similar results for the other pairs).  We measure $\reszz$ using a modified echo sequence~\cite{Negirneac21} and $\jone$ ($\jtwo$) by spectroscopy of the $\ket{0_1 1_0}-\ket{1_1 0_0}$  $\left(\ket{1_1 1_0}-\ket{0_1 2_0}\right)$ avoided crossing reached by flux biasing $\qc$ ($\qa$) away from sweetspot. Decoherence limits the ability of both methods to accurately extract these quantities at their zero crossings. When necessary, we rely on simple polynomial fits to exact the $\cca$ frequencies at which they are nulled. The results clearly illustrate that the three residual interaction strengths can be nulled, but not simultaneously.

We first focus on single-qubit gates, studying the impact of $\cca$ bias frequency on the individual and simultaneous single-qubit gate performance  of $\qa$ and $\qc$ [Fig.~\ref{fig:SingleQubitGates}(b)]. Individual single-qubit randomized benchmarking (RB) with the other transmon kept in $\ket{0}$ provides a baseline single-qubit native gate error of $\sim 0.2\%$ across most of the $650~\MHz$ range covered in $\cca$. We attribute the increase observed at the lower end of $\cca$ frequency to increased transmon dephasing due to hybridization with the coupler and increased sensitivity of the bare coupler to $1/f$ flux noise~\cite{SOM_tc_spect}. Simultaneous performance matches individual performance over a wide $\cca$ range, including $\reszz^{01}=0$ and $\jone^{01}=0$. A degradation is observed in simultaneous performance by $\jtwo=0$. Note that the $\cca$ sweetspot limits our ability to extensively cover the $\reszz^{01}>0$ regime. This is intentional: we strategically targeted couplers to have sweetspots just above $\reszz^{01}=0$ to avoid unnecessarily comprising on their coherence.

During this above study, we observed that $\reszz^{01}=0$ is not the optimal choice for readout. Specifically, at this $\cca$ bias, we found it impossible to achieve an average assignment error~\cite{Chen22} $\epsro<2.5\%$ for $\qc$, whose qubit transition frequency is above that of $\qa$. The underlying physical reason is population exchange $\ket{1_0 0_1}\rightarrow \ket{0_0 0_1}$ activated during readout: photons injected to the $\qc$ readout mode cause a Stark shift of $\qc$, displacing its qubit transition downward in frequency and broadening its linewidth~\cite{Schuster05}, causing spectral overlap with $\qa$. Similar physics has been used to explain enhanced qubit relaxation during readout, caused by Stark-shift activated exchange with lower-frequency two-level defects~\cite{Thorbeck24}. Figure~\ref{fig:J1ROPExch} evidences the phenomenon and offers a mitigation strategy, showing the loss of excitation from $\qc$ to $\qa$ as a function of both $\qc$ readout pulse amplitude and $\cca$ bias. Population exchange is strongly suppressed in a $\cca$ frequency range centered around $\jone^{01}=0$, where $\epsro<1\%$ is reached for $\qc$ with a total readout duration of $800~\ns$ (including photon depletion). As all coupled transmon pairs have $\reszz/(2\pi)<100~\kHz$ at $\jone=0$, and in consideration of Fig.~~\ref{fig:SingleQubitGates}, it seems tempting to choose this bias as the default for all couplers. However, it remains important to first study spectator effects on two-qubit gates before reaching conclusions.

\begin{figure}
\centering
\includegraphics{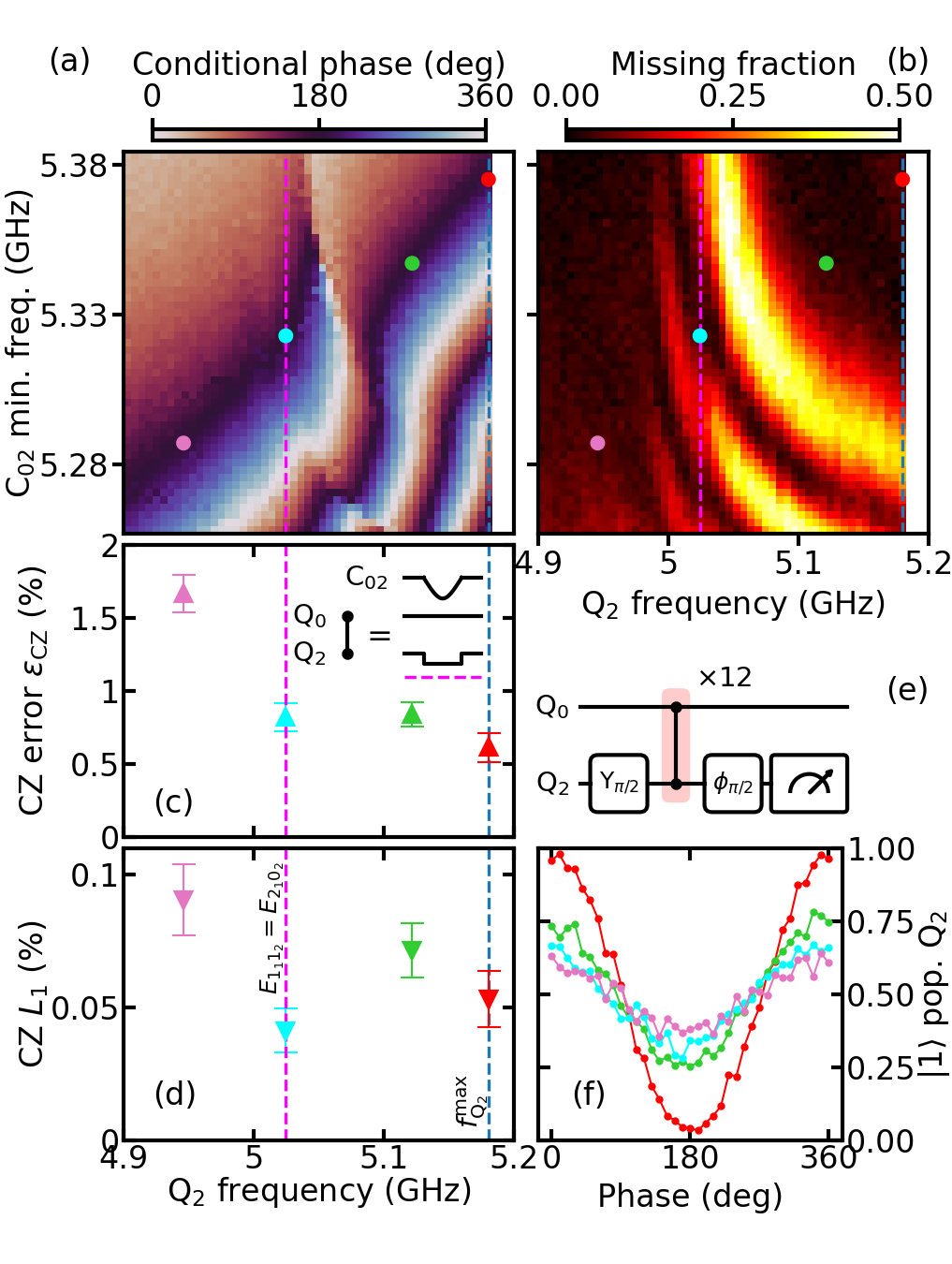}
\caption{
Study of a CZ gate between $\qb$ and $\qc$, implemented by applying a fast-adiabatic flux pulse to coupler $\ccb$ and a square pulse to the lower-frequency qubit, $\qb$. (a,b) Conditional phase and missing fraction extracted from conditional oscillation experiments. (c,d) CZ error and leakage measured using interleaved RB at four points along the contour of $180^\degrees$ conditional phase [colored markers in (a,b)]. Error bars are the standard deviation of the mean computed from 5 repetitions. The inset in (c) is a cartoon representation of frequency excursions of $\ccb$, $\qc$ and $\qb$. (e) Schematic of a simple experiment to study the impact that flux pulsing $\qb$ during CZ gates has on its coherence. We embed 12 CZ gates in a Ramsey sequence on $\qb$.  (f) Ramsey oscillations for the four chosen CZ settings, clearly showing that the impact on coherence is lowest when $\qb$ remains at its sweetspot, as expected.
\label{fig:CZLandscape}}
\end{figure}

\begin{figure}
\centering
\includegraphics{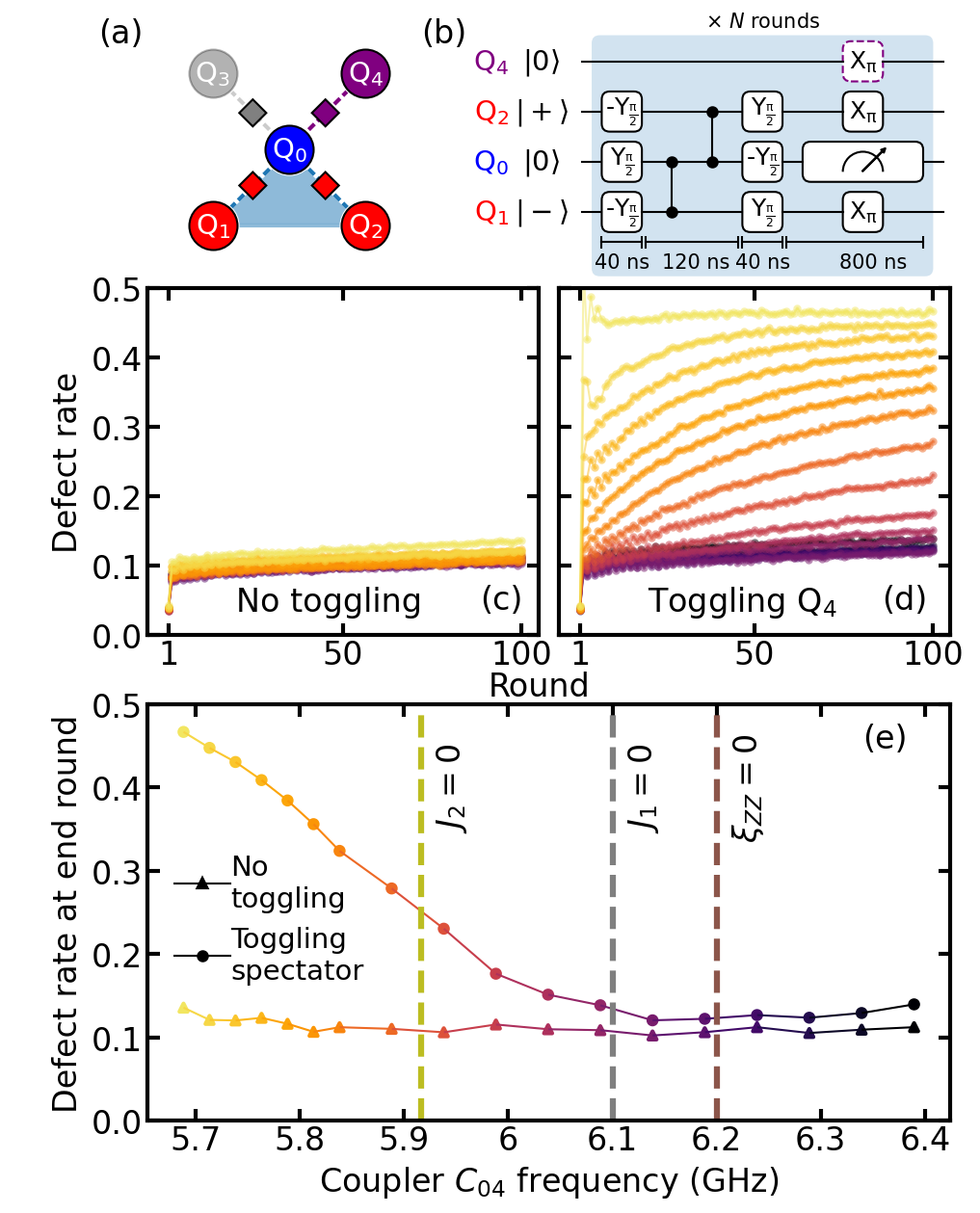}
\caption{
Study of spectator effects during repeated parity checks. (a) Simplified schematic of the device with color coding to identify data qubits ($\qa$ and $\qb$, red), ancilla ($\qc$, blue) and spectator ($\qe$, purple). (b) Quantum circuit realizing the indirect measurement of $X_2 X_1$ parity using $\qc$ as ancilla. (c) Defect rates measured over 100 rounds at different $\cce$ frequencies (represented by the color scale), keeping spectator $\qe$ in the ground state. (d) Defect rates measured now toggling $\qe$ at every round. (e) Common plot of the defect rate at round 100 for data in (c) and (d). Vertical lines mark the $\reszz^{04}=0$, $\jone^{04}=0$ and $\jtwo^{04}=0$ coupler bias points.
\label{fig:SpectDefRate}}
\end{figure}

To this end, we must now choose an implementation of CZ gates drawing inspiration from prior work on similar tunable-coupling architectures~\cite{Marxer23, Sung21b, Stehlik21, Collodo20, Xu20, Sete21} (Fig.~\ref{fig:CZLandscape}). A general controlled-phase gate can be implemented by dynamically detuning the coupler, controllably increasing $\reszz$ until the target conditional phase is acquired~\cite{Stehlik21,Collodo20,Xu20}. We study the conditional phase and missing fraction (a proxy for leakage)~\cite{Rol19} of pair $(\qb,\qc)$ in response to a unipolar, fast-adiabatic pulse~\cite{Martinis14} on $\ccb$ and a square pulse to the lower-frequency transmon $\qb$, both of duration $60~\ns$. The conditional-phase landscape displays a contour of $180^\degrees$, the required phase for CZ. We select four points along this contour to evaluate the performance of CZ gates using interleaved RB with modifications to quantify leakage~\cite{Magesan12,Wood18}. The best performance is achieved when $\qb$ remains at sweetspot (i.e., only the coupler is pulsed), where we extract a CZ gate error $\epscz=(0.62\pm0.03)\%$ with leakage $\lone=(0.051\pm0.014)\%$. Most likely, this choice maximizes fidelity because keeping $\qb$ at sweetspot maximizes its coherence, as shown in Figs.~\ref{fig:CZLandscape}(e-f). However, we did not perform detailed numerical simulation to dissect the contribution of different physical error sources to the error budget~\cite{Rol19, Negirneac21}.

We can now study the combined effect of spectators on readout, single- and two-qubit gates in a minimal setting relevant for QEC~\cite{Fowler12}. Specifically, we perform repeated indirect measurement of the parity operator $X_2 X_1$ using $\qc$ as ancilla and $\qe$ as spectator (Fig.~\ref{fig:SpectDefRate}). We extract the defect rate up to 100 rounds as a function of $\cce$ bias frequency without and with toggling spectator $\qe$ in every round 
(similar results are obtained when using $\qd$ as the spectator~\cite{SOM_tc_spect}.) The defect rate at a given round is defined as the fraction of time that the declared parity outcome does not match that declared in the previous round.  When $\qe$ is not toggled, the defect rate is largely insensitive to $\cce$ positioning, remaining $<12\%$ over all bias positions and all rounds. In contrast, when $\qe$ is toggled, the defect rate remains low and nearly round independent only near $\reszz^{04}=0$. A slight degradation is observed by $\jone^{04}=0$ (more noticeable when toggling $\qd$ instead of $\qe$~\cite{SOM_tc_spect}). As $|\reszz^{04}|$ further increases, the defect rates rise, showing a positive slope toward a higher steady state with increasing rounds. This slope is a hallmark of leakage~\cite{Google21,McEwen21} to higher-excited transmon states (of $\qc$ in this case). Note that because the $\qc$ qubit frequency is below that of $\qe$, no measurement-induced exchange is expected to take place in the one-excitation manifold. However, exchange can occur in the two-excitation manifold as $\ket{1_4 1_0}$ is Stark-shifted toward $\ket{2_4, 0_0}$ (In \cite{SOM_tc_spect} we show that this effect is suppressed at the $\jtwo^{04}=0$ bias). This effect could underlie the small increase in defect rate at $\reszz=0$ observed when toggling $\qe$. 

In summary, we have performed an experimental study to identify the best frequency positioning of flux-tunable, transmon-like couplers to mitigate spectator effects on quantum operations in a circuit QED processor. We show that this type of tunable coupler can null $ZZ$ and exchange couplings, but not simultaneously. For single-qubit gates, spectator effects are minimized at a coupler bias region that spans $\jone=0$ and $\reszz=0$. For two-qubit gates, the spectator defect rate study suggests that $\reszz=0$ is the preferable bias configuration. For readout, however, spectator effects are only minimized at $\jone=0$ ($\jtwo=0$) when the measured transmon is above (below) the coupled spectator. Consequently,
it could prove beneficial to dynamically switch between nulling conditions for different quantum operations, taking advantage of the fast flux tunability already in place and calibrated for two-qubit gates. Moving forward, it will also be important to further test our device architecture and spectator-effect mitigation strategies on larger devices, for example the 17-transmon, 24-coupler processor required for the distance-3 surface code~\cite{Krinner22, Google23, Besedin25}, where all transmons couple to either two, three, or four nearest neighbors. It is also interesting to explore alternative tunable-coupling strategies with the capability to simultaneously null $ZZ$ and exchange interactions~\cite{Heunisch23}.

\section{Data availability}
The data for all figures in main text and supplement are available online at
\verb"http://github.com/DiCarloLab-Delft/"\\
\verb"SpectatorEffectsTunableCouplers".
Requests for additional materials should be addressed to LDC (l.dicarlo@tudelft.nl).

\section{Acknowledgements}
We thank S.~Dobrovolskiy and G.~Limodio  of QuantWare for fabricating the device,  F.~Battistel, A.~Radhakrishnan and R.~Navarathna of Qblox for assistance with electronics, H.~Ali for insightful discussions, and V.~Sinha for project management assistance. We are grateful to G.~Calusine and W.~Oliver for providing the traveling-wave parametric amplifier used in the amplification chain. This research is funded by the Netherlands Organization for Scientific Research (NWA.1292.19.194), Quantum Delta NL (DiagnostiQ and National Growth Fund KAT-1), the Allowance for Top Consortia for Knowledge and Innovation (TKI) of the Dutch Ministry of Economic Affairs, the European Union Flagship on Quantum Technology (OpenSuperQplus100, No. 101113946), and Intel Corporation.

\section{Author contributions}
SVS and LDC conceptualized the experimental study. SVS. performed the experiment with contributions from THFV, TRA, AL and MAR. SVS performed the data analysis with contributions from THFV, SLMM and YX. SVS, THFV and LDC designed the Hamiltonian parameters of device, with contributions from VS. TS and FB designed the device layout to target these parameters. SVS and LDC wrote the manuscript with contributions from TS and FB, with feedback from all coauthors. LDC supervised the project.

\section{Conflicts of interests}
The authors declare no conflict of interests.

\bibliographystyle{apsrev4-2}

\end{bibunit}
\clearpage

\renewcommand{\theequation}{S\arabic{equation}}
\renewcommand{\thefigure}{S\arabic{figure}}
\renewcommand{\thetable}{S\arabic{table}}
\renewcommand{\bibnumfmt}[1]{[S#1]}
\renewcommand{\citenumfont}[1]{S#1}
\setcounter{figure}{0}
\setcounter{equation}{0}
\setcounter{table}{0}

\begin{bibunit}[apsrev4-2]

\onecolumngrid
\section*{Supplementary material for ``\nametitle''}
\FloatBarrier
\twocolumngrid

This supplementary material provides additional information supporting the claims and figures of the main text.

\section*{Experimental setup}

\begin{figure}
\centering
\includegraphics[width=\columnwidth]{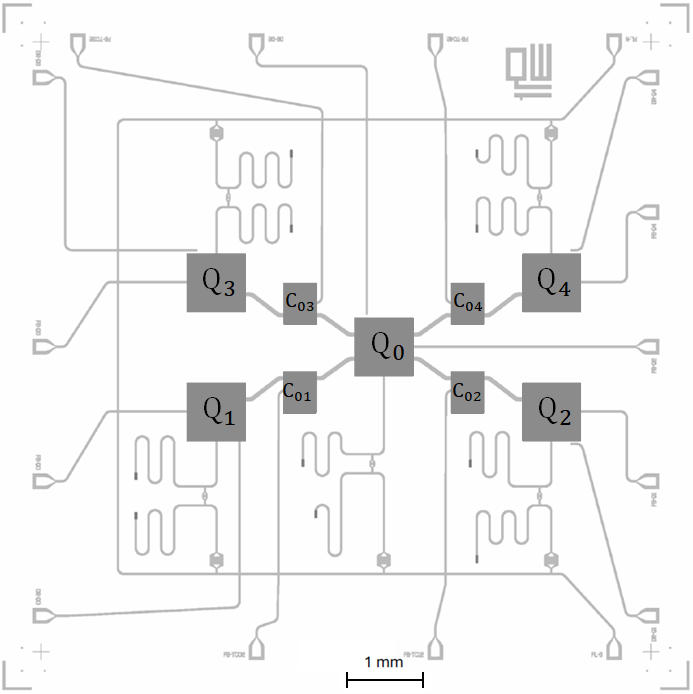}
\caption{Schematic of the 5 transmon, 4 tunable coupler device used in this experiment.
\label{fig:device_layout}}
\end{figure}

The device used consists of 5 transmons in a starfish configuration, with 4 dedicated tunable couplers connecting the central transmon $\qc$ to each of the 4 corner ones (Fig.~\ref{fig:device_layout}). Every transmon has a dedicated microwave-drive line and flux-control line, as well as a dedicated pair of readout Purcell-filter resonators~\cite{Heinsoo18} for independent readout via one common feedline. Tunable couplers have dedicated flux-control lines. The otherwise planar device, fabricated from a Ta base layer sputtered on a Si substrate, has Al crossovers and airbridges, including 'shoelacing' airbridges enabling post-fabrication frequency trimming of readout resonator and Purcell filters~\cite{Valles23}.

The device is cooled in a Leiden Cryogenics CS81 dilution refrigerator with $\sim10~\mK$ base temperature, fitted with standard attenuation and filtering in all input lines, and amplification in the readout chain provided by a travelling-wave parametric amplifier~\cite{Macklin15} (at the mixing chamber) and a high-electron-mobility-transistor amplifier at $4~\K$. All control and readout signals are generated and acquired at room temperature using a Qblox Cluster that contains one QRM-RF module for readout, three QCM-RF modules for single-qubit gates and three QCM modules for baseband flux control of transmons and couplers. All the experiments are scheduled using the open-source framework Quantify, together with the Superconducting Qubit Tools package of Orange Quantum Systems.

\begin{figure*}
\centering
\includegraphics{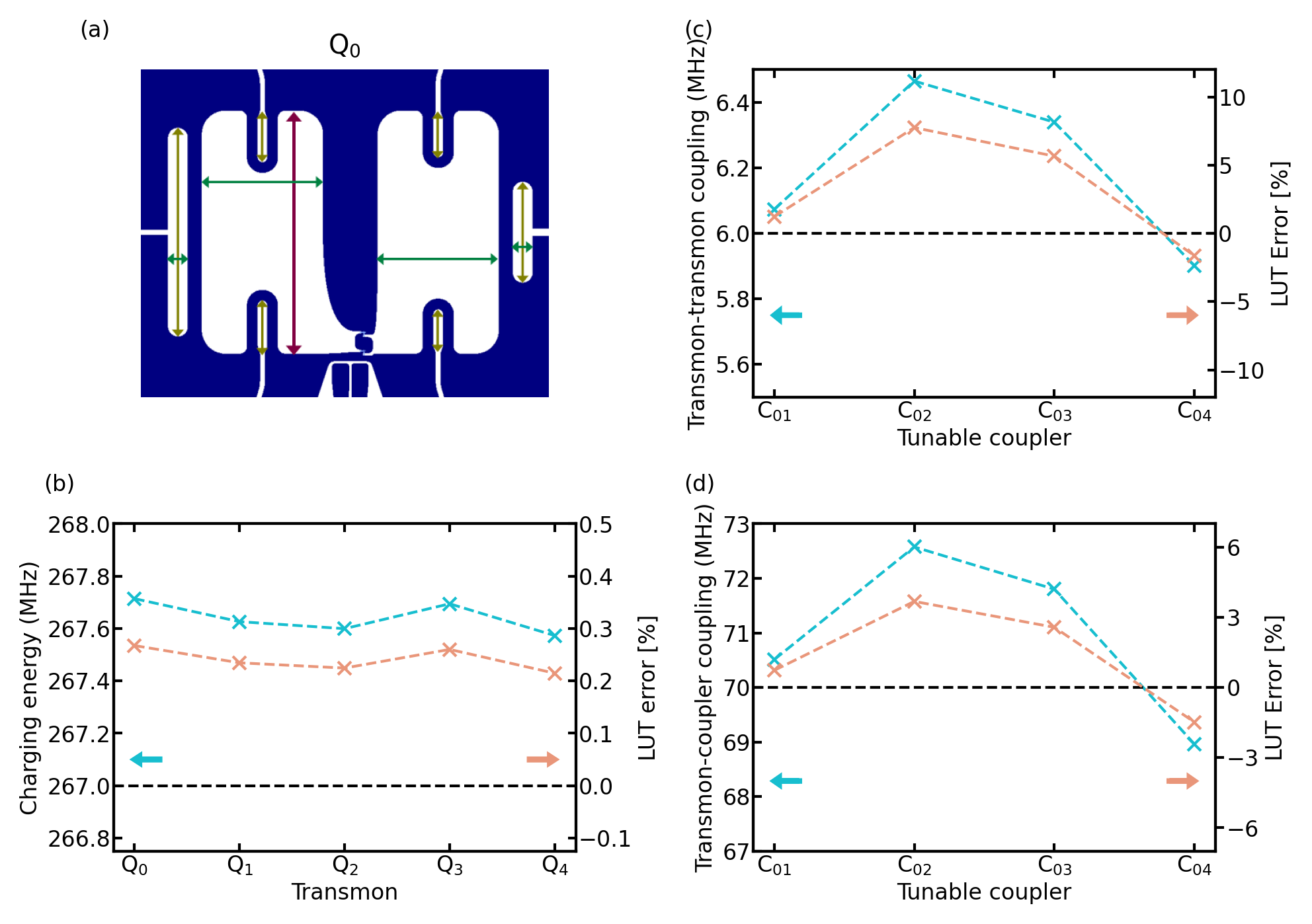}
\caption{Fine-tuning of the transmon and tunable-coupler capacitances using the lookup-table method. (a) Schematic of the transmon geometry, showing the 7 adjustable dimensions that are tuned for capacitance targeting. The six gold-colored arrows indicate dimensions used for coupling capacitances. The dimensions of the main pads (red arrow) are used to set the total transmon capacitance. Other dimensions (green arrows) are kept constant during tunable-coupler design. (b) Charging energy, (c) qubit-qubit coupling strength and (d) qubit-coupler coupling strength as predicted by the LUT method, and prediction error determined with full finite-element simulation.
\label{fig:lookup_table_tuning}}
\end{figure*}

\begin{figure*}
\centering
\includegraphics{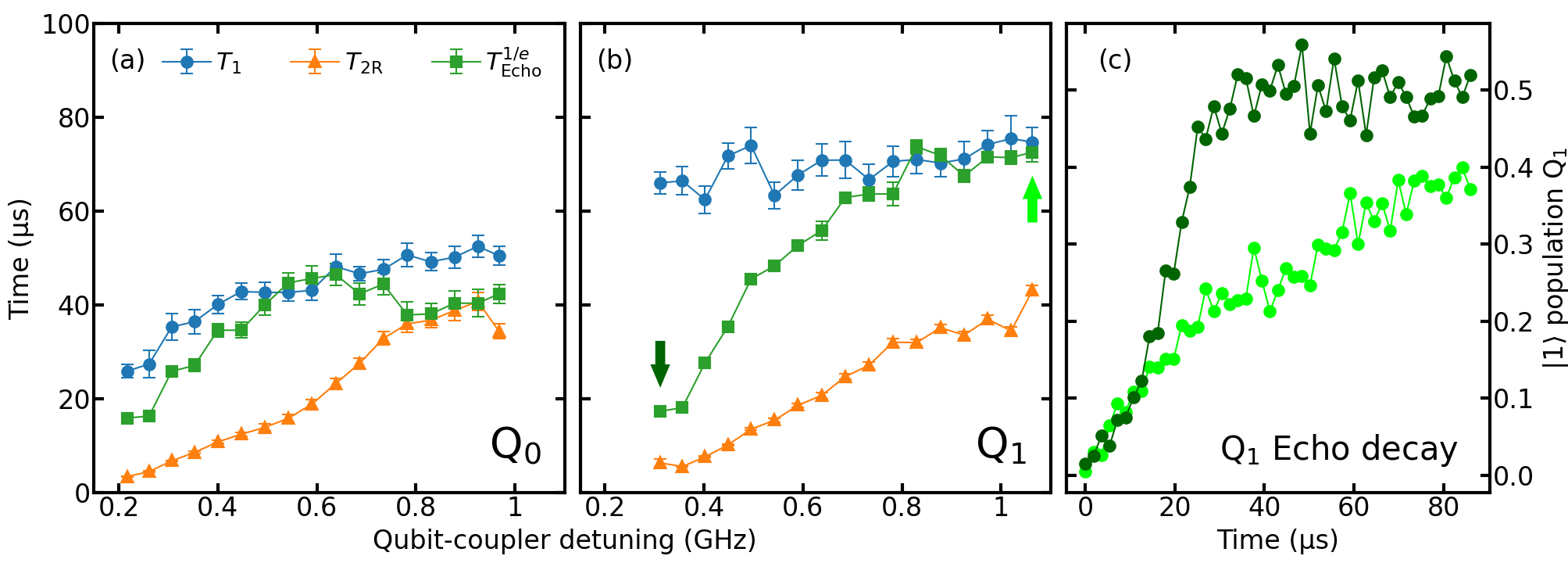}

\caption{Relaxation and dephasing (Ramsey and echo) times of transmons (a) $\qc$ and (b) $\qa$ at their sweetspot, as a function of coupler $\cca$ frequency detuning. Error bars are the standard deviation of the mean computed from 35 repetitions. As the coupler is biased away from its sweetspot and toward the transmons, the increasing sensitivity of the bare coupler to flux noise and the increasing hybridization affect the transmon dephasing times. Relaxation time is also worsened by the increased hybridization as the bare coupler has lower relaxation time than the transmons. (c) Echo decay of $\qa$ for two coupler biases: (light green) coupler at sweetspot, (dark green) coupler just $\sim300~\MHz$ above $\qa$. Note the transition from exponential to gaussian decay resulting from the increase in sensitivity of the bare coupler to $1/f$ flux noise~\cite{Braumuller20} and the increase in transmon-coupler hybridization.
\label{fig:qubit_coherence_coupler}}
\end{figure*}

\begin{table*}
    \centering
    \begin{tabular}{|c|c|c|c|c|c|}
        \hline
         & $\qc$ & $\qa$ & $\qb$ & $\qd$ & $\qe$ \\ \hline \hline
        Sweetspot frequency (GHz) & $5.295$ & $5.218$ & $5.181$ & $5.463$ & $5.457$ \\ \hline
        Anharmonicity (MHz) & $-275$ & $-285$ & $-286$ & $-279$ & $-281$ \\ \hline
        $T_1$ ($\us$) & $31.5\pm0.8$ & $76.2\pm1.3$ & $68.2\pm0.8$ & $81.4\pm1.2$ & $34.1\pm0.4$ \\ \hline
        $T_{2\mathrm{R}}$ ($\us$) & $12.0\pm0.2$ & $29.2\pm1.1$ & $33.7\pm0.8$ & $34.1\pm0.4$ & $10.2\pm0.9$ \\ \hline
        $T_{2\mathrm{E}}$ ($\us$) & $32.8\pm0.6$ & $70.0\pm2.2$ & $95.9\pm1.3$ & $80.0\pm1.2$ & $35.7\pm2.4$ \\ \hline
        Single-qubit gate error ($\%$) & $0.056\pm0.015$ & $0.08\pm0.02$ & $0.023\pm0.002$ & $0.057\pm0.016$ & $0.043\pm0.001$ \\ \hline
    \end{tabular}
    \caption{Summary of key qubit parameters. The relaxation and coherence times are measured with all the couplers at  $\jone=0$ bias. The uncertainties represent the standard deviation of the mean over repeated measurements.}
    \label{tab:qubit_summary_table}
\end{table*}

\begin{table*}
    \centering
    \begin{tabular}{|c|c|c|c|c|}
        \hline
         & $\cca$ & $\ccb$ & $\ccd$ & $\cce$ \\ \hline \hline
        Sweetspot frequency (GHz) & $6.275$ & $6.296$ & $6.317$ & $6.383$ \\ \hline
        Anharmonicity (MHz) & $-250$ & $-245$ & $-253$ & $-255$ \\ \hline
        $T_1$ at sweetspot ($\us$) & $37.4\pm8.0$ & $36.1\pm6.0$ & $27.5\pm9.0$ & $50.1\pm3.5$ \\ \hline
        $T_{2\mathrm{R}}$ at sweetspot ($\us$) & $6.8\pm2.5$ & $16.6\pm1.6$ & $14.1\pm3.9$ & $21.2\pm3.0$ \\ \hline
        $T_{2\mathrm{E}}$ at sweetspot ($\us$) & $14.0\pm2.9$ & $34.6\pm4.9$ & $16.7\pm3.3$ & $26.9\pm3.0$ \\ \hline
        CZ gate error ($\%$) & $0.49\pm0.33$ & $0.61\pm0.10$ & $0.72\pm0.05$ & $0.88\pm0.07$ \\ \hline
        CZ leakage ($\%$) & $0.005\pm0.024$ & $0.053\pm0.011$ & $0.075\pm0.008$ & $0.005\pm0.014$ \\ \hline
    \end{tabular}
    \caption{Summary of key tunable coupler parameters. The relaxation and coherence times are measured with each coupler individually at the sweetspot, while the other couplers are placed at $0.5\Phi_0$ flux. The CZ gates are calibrated and characterized with all the tunable couplers biased $\jone=0$.}
    \label{tab:coupler_summary_table}
\end{table*}

\begin{figure}
\centering
\includegraphics{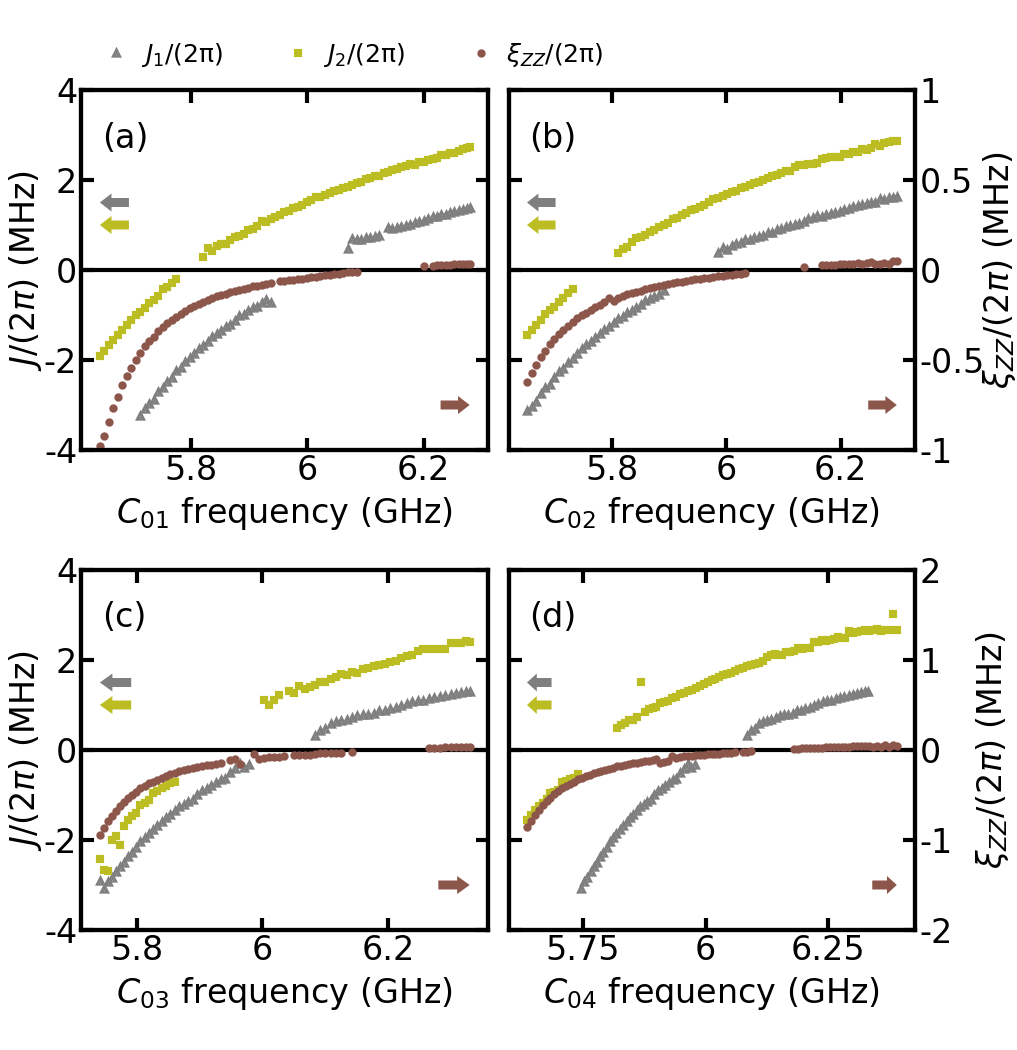}
\caption{$\jone^{0i}$, $\jtwo^{0i}$ and $\reszz^{0i}$ for all transmon pairs $\left(\qc,\qi\right)$ as a function of the frequency of the coupler $\cci$ connecting them: (a) $\cca$, (b) $\ccb$, (c) $\ccd$, (d) $\cce$.
\label{fig:ZZJ1J2allpairs}}
\end{figure}

\begin{figure}
\centering
\includegraphics{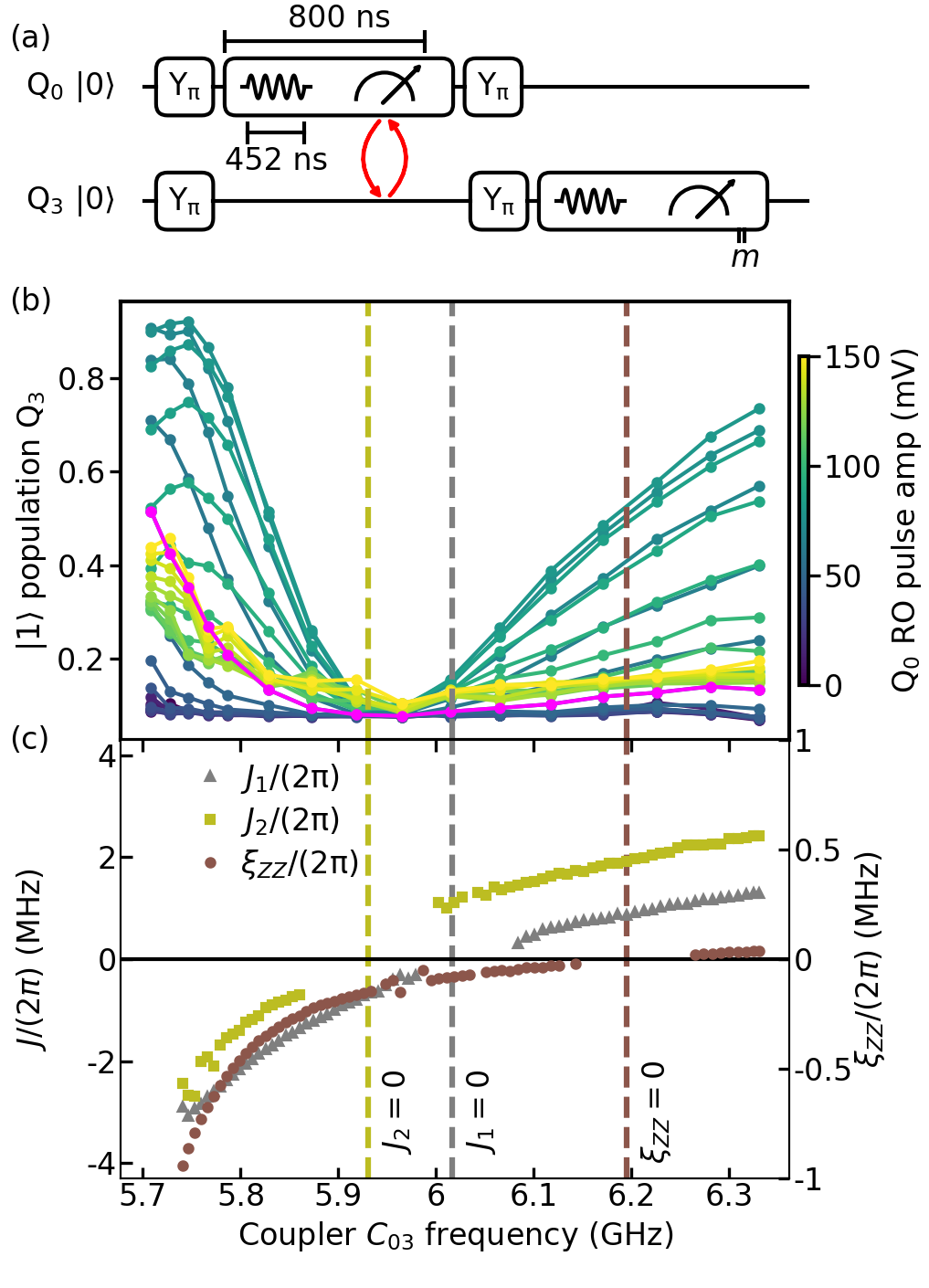}
\caption{Measurement-induced population exchange in the two-excitation manifold. (a) Circuit schematic of the experiment implemented on tranmons $\qc$ and $\qd$. (b) Population of $\qd$ second-excited state, measured as a function of $\ccd$ frequency and amplitude of the $\qc$ readout pulse. (c) $\jone^{03}$, $\jtwo^{03}$ and $\reszz^{03}$ as a function of $\ccd$ frequency [same data as Fig.~\ref{fig:ZZJ1J2allpairs}(c)].
\label{fig:J2ROPExch}}
\end{figure}

\begin{figure}
\centering
\includegraphics{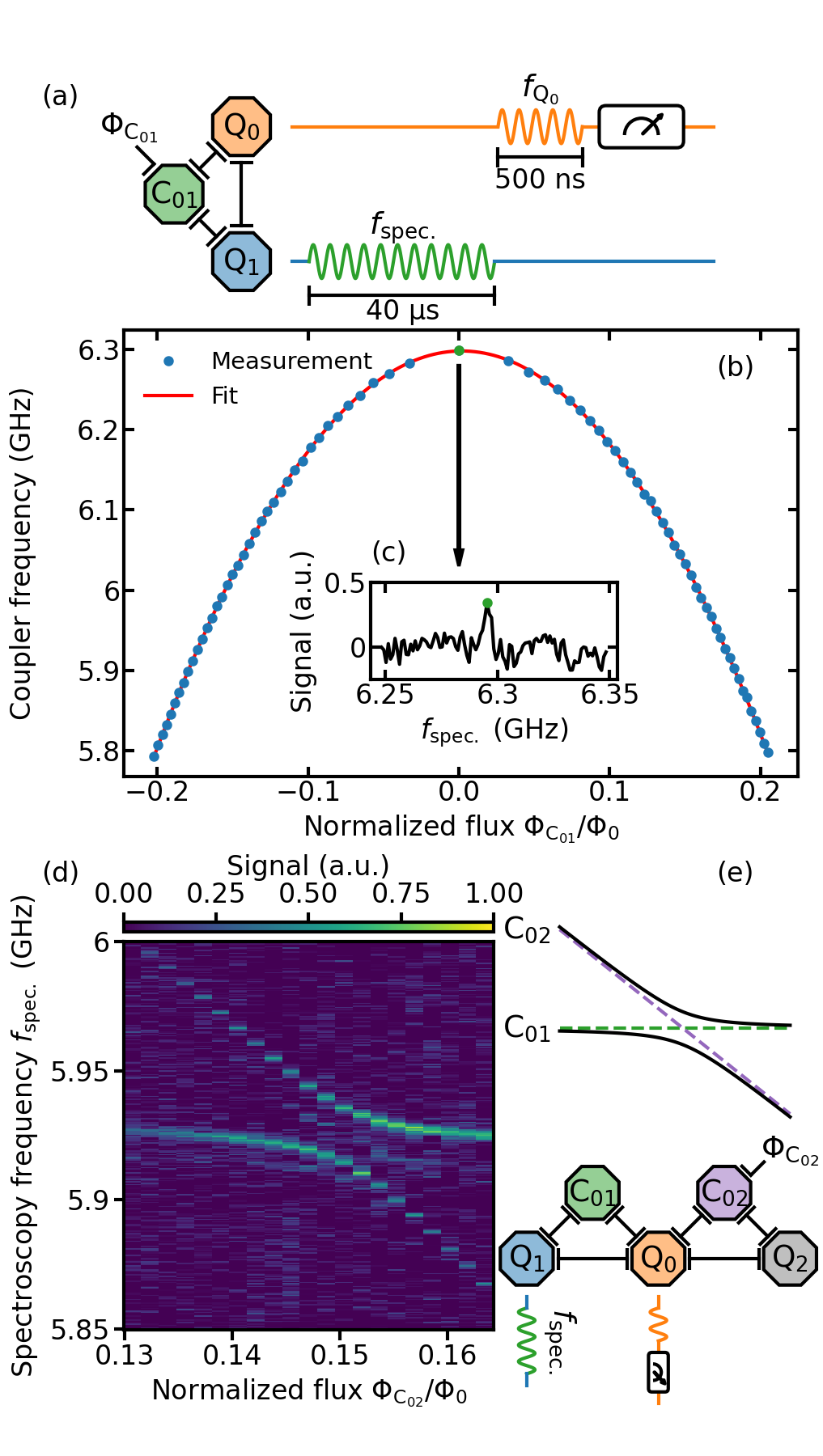}
\caption{(a) Pulsed spectroscopy sequence on coupler $\cca$. A spectroscopic pulse is applied to the drive line of $\qa$ and $\qc$ then readout. (b) $\cca$ flux arc obtained by this spectroscopy method. A polynomial best fit is used to convert from static flux to $\cca$ frequency. Inset (c): example single spectroscopy trace, revealing the frequency of $\cca$ at its flux insensitive point (sweetspot). (d) Spectroscopy of $\cca$ while sweeping the flux bias of $\ccb$ in a region where the two elements hybridize, producing an avoided crossing with minimum splitting of $17~\MHz$. (e) Schematic of the system of two interacting tunable couplers.
\label{fig:CouplerSpectroscopy}}
\end{figure}

\begin{figure}
\centering
\includegraphics{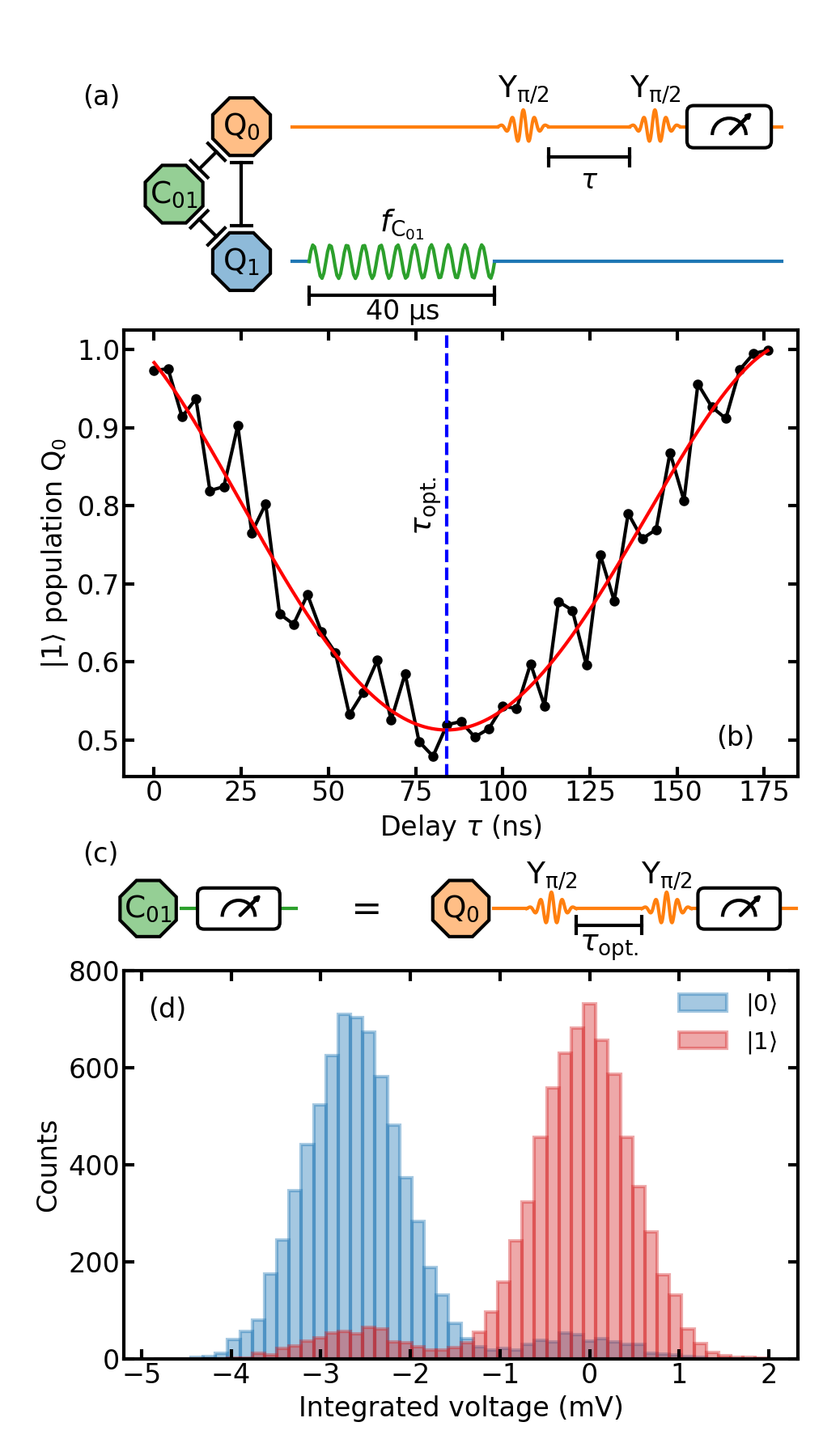}
\caption{Indirect readout of coupler $\cca$ via $\qc$. (a) Calibration using a long spectroscopy pulse on $\cca$ (applied on the microwave drive line of $\qa$) and a Ramsey sequence on $\qc$. (b) Excited-state population of $\qc$, as a function of delay $\tau$ between the two $\pihalf$ pulses on $\qc$. (c) Optimized quantum circuit to measure $\cca$ via $\qc$. (d) SSRO characterization for $\cca$, showing $\epsro=8\%$.
\label{fig:CouplerReadout}}
\end{figure}

\begin{figure}
\centering
\includegraphics{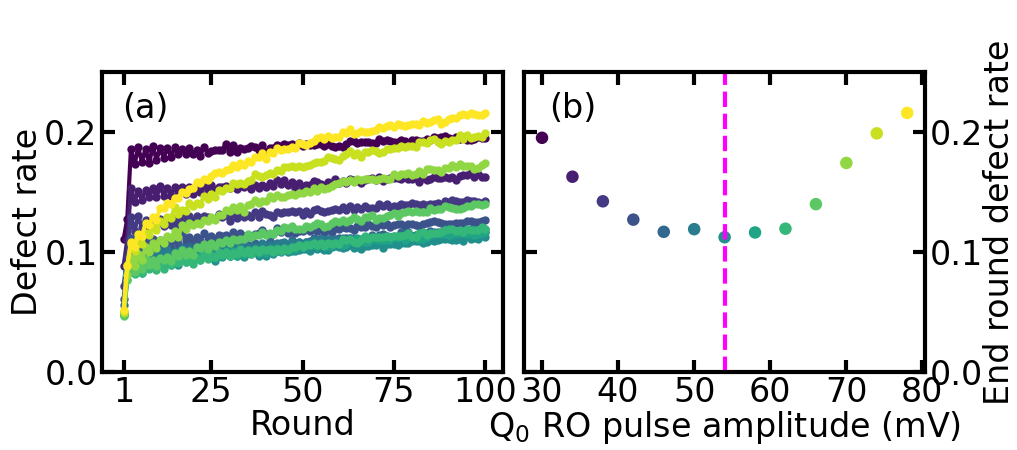}
\caption{Parity-check defect rates for different amplitudes of the ancilla $\qc$ readout pulse. (a) Defect rates measured over 100 rounds. (b) Defect rate at round 100 as a function of $\qc$ readout pulse amplitude. We choose the readout pulse amplitude (pink vertical line) that minimizes the defect rate at round 100.
\label{fig:DefRateROPulseAmp}}
\end{figure}

\begin{figure}
\centering
\includegraphics{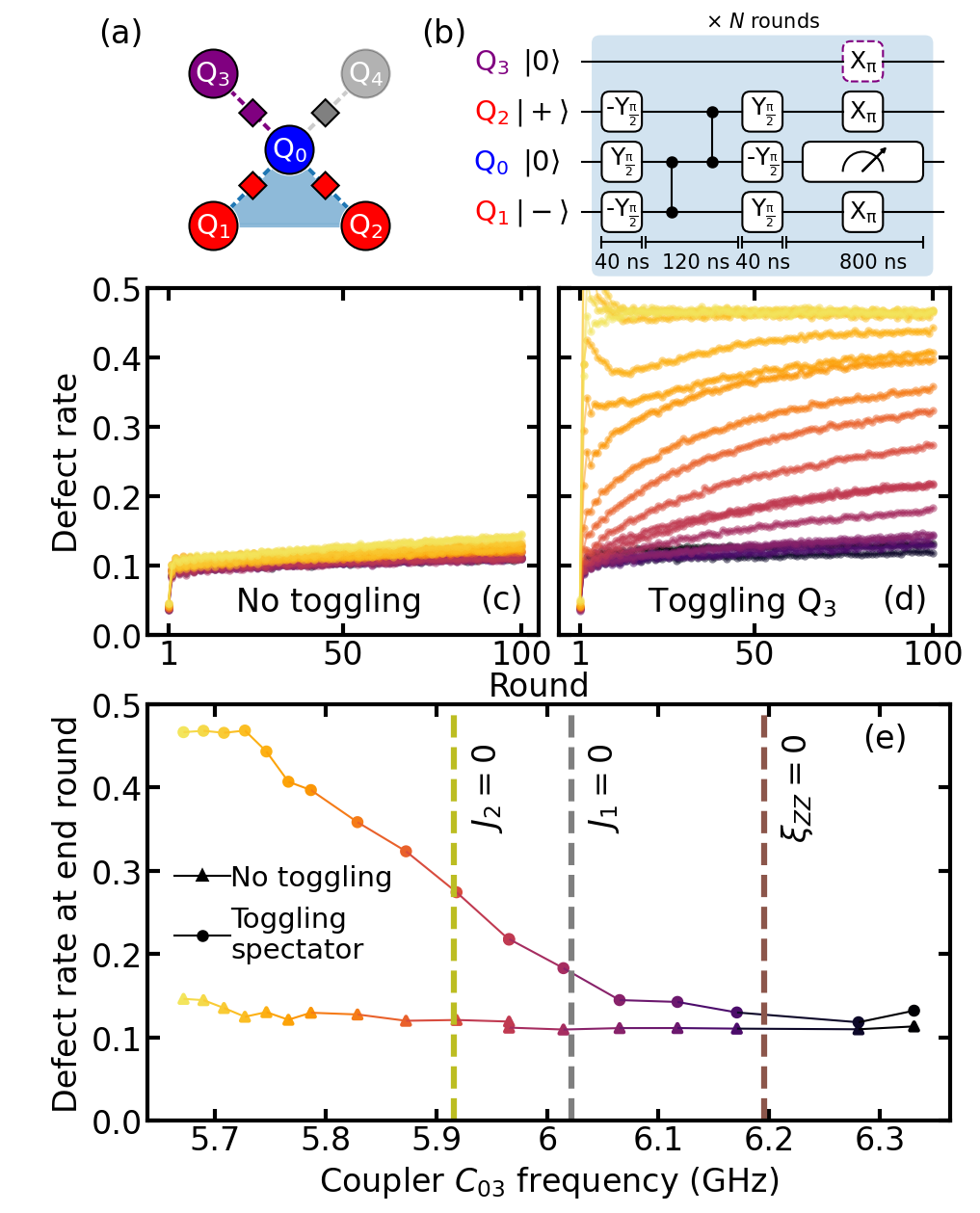}
\caption{Same experiment as in Fig.~4, but using  $\qd$ instead of $\qe$ as the spectator. (a) Simplified schematic of the device with color coding to identify data qubits ($\qa$ and $\qb$, red), ancilla ($\qc$, blue) and spectator ($\qd$, purple). (b) Quantum circuit realizing the indirect measurement of $X_2 X_1$ parity using $\qc$ as ancilla.
(c) Defect rates measured over 100 rounds at different $\cce$ frequencies (represented by the color scale), keeping spectator $\qd$ in the ground state. (d) Defect rates measured now toggling $\qd$ at every round. (e) Common plot of the defect rate at round 100 for data in (c) and (d). Vertical lines identify the $\ccd$ frequencies at which $\reszz=0$ and $\jone=0$. Vertical lines mark the $\reszz^{03}=0$, $\jone^{03}=0$ and $\jtwo^{03}=0$ coupler bias points.
\label{fig:SpectDefRateSOM}}
\end{figure}

\begin{figure}
\centering
\includegraphics{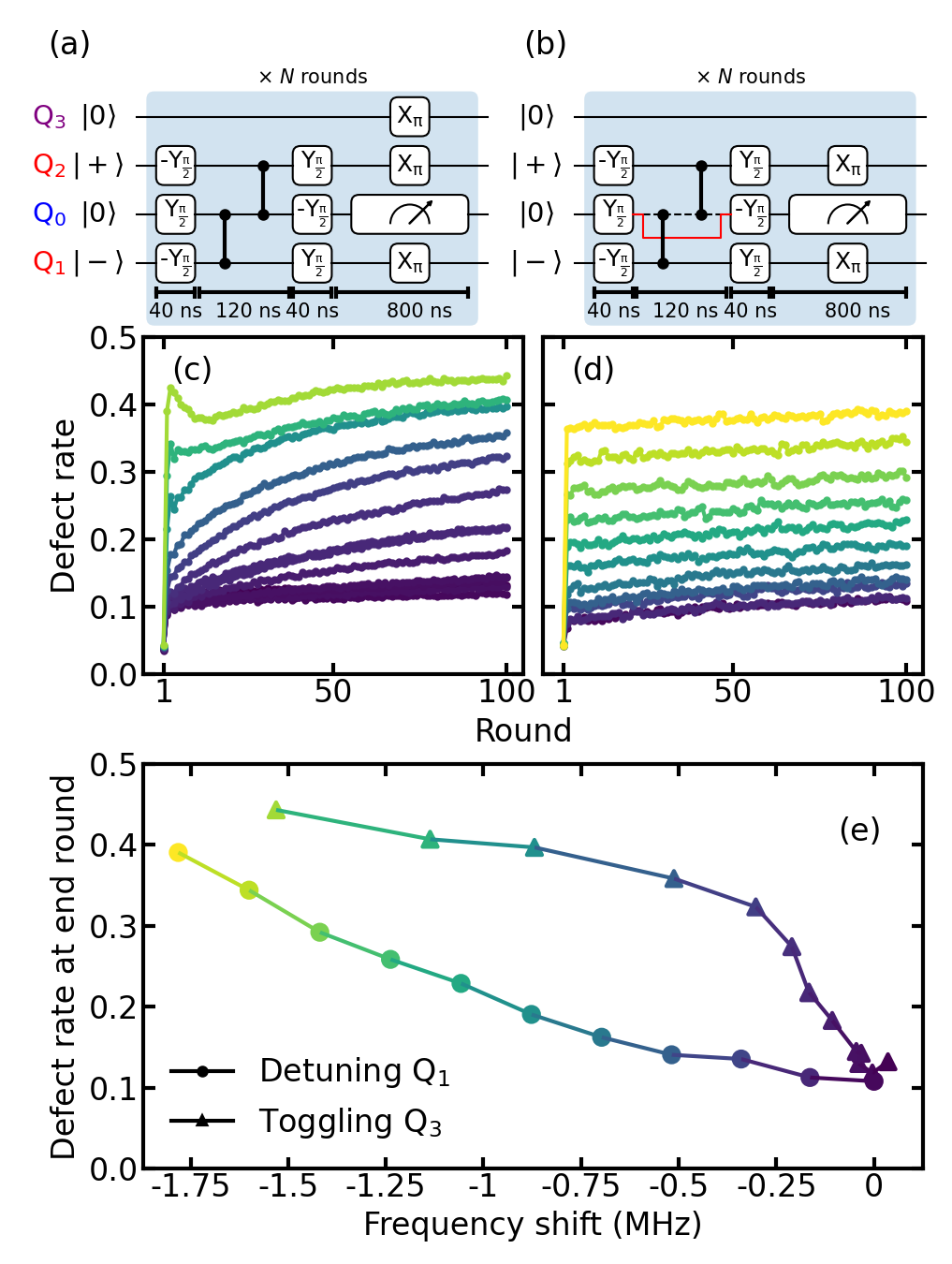}
\caption{Emulating residual $ZZ$ coupling errors by flux pulsing $\qc$. (a,c) Same experiment and data as in Fig.~\ref{fig:SpectDefRateSOM}. (b) Weight-2 $X$-type parity check experiment, marking that $\qc$ is now dynamically detuned. (d) Defect rates measured for different frequency shift for $\qc$. (e) Comparison of defect rates at round 100 for both cases.
\label{fig:PulsedDefectRate}}
\end{figure}

\section*{Tunable-coupler design}
As discussed in the main text, our target exchange coupling strengths for the system consisting of two transmons connected by a tunable coupler are $\gqq=2\pi \times 6~\MHz$ and $\gqc=2\pi \times 70~\MHz$. Translating these couplings into geometry requires fine tuning of the capacitive couplings. Approximately~\cite{Sete21},
\begin{equation*}
\gqq = \frac{E_{ij}}{\sqrt2}\left(\frac{E_{\mathrm{J}_i}}{E_{\mathrm{C}_i}}\frac{E_{\mathrm{J}_j}}{E_{\mathrm{C}_j}}\right)^\frac{1}{4},
\end{equation*}
\noindent
where $E_{\mathrm{J}_i}$ and $E_{\mathrm{J}_j}$ are the Josephson energies of the two coupled trasmons $\mathrm{Q}_i$ and $\mathrm{Q}_j$, while $E_{\mathrm{C}_i}$ and $E_{\mathrm{C}_j}$ are their charging energies. $E_{ij}$ represents the direct capacitive coupling energy between both transmons. The transmon parameters ($E_\mathrm{J}$ and $E_\mathrm{C}$) are previously fixed while designing the layout, leaving only $E_{ij}$ as an adjustable parameter used to target $\gqq$.

% which only depends on (fixed) qubit parameters and the direct capacitive coupling $E_{\mathrm{12}}$. Here, $E_J$ is the Josephson energy and $E_C$ is the charging energy. We tune the direct qubit-qubit coupling by adjusting the capacitance of the pad connecting the qubits to the coupler until the desired value is reached.

Similarly~\cite{Sete21},
\begin{equation*}
        \gqc = \frac{E_{i\mathrm{c}}}{\sqrt2}\left(\frac{E_{\mathrm{J}i}}{E_{\mathrm{C}i}}\frac{E_{\mathrm{Jc}}}{E_{\mathrm{Cc}}}\right)^\frac{1}{4},
\end{equation*}
\noindent
where $E_{\mathrm{Jc}}$ and $E_{\mathrm{Cc}}$ represent the Josephson and charging energies of the tunable coupler and $E_{i\mathrm{c}}$ the coupling energy between transmon $\mathrm{Q}_{i}$ and the coupler. In our architecture, $E_{\mathrm{ic}}$ is dependent on $E_{ij}$, since a single coupling pad is used to couple all three elements, and it is therefore fixed when targeting $\gqq$. The value of $E_{\mathrm{Jc}}$ is constrained to target the coupler sweetspot frequency $100~\MHz$ above the $\reszz=0$ point, leaving $E_{\mathrm{Cc}}$ as the only adjustable parameter when targeting $\gqc$.

\subsection*{Fast calculation of the capacitance matrix}

The methods described above allow coarse tuneup of the exchange coupling stengths. In a more realistic scenario, some of the design parameters have non-negligible impact on each other. For example, decreasing $E_{\mathrm{qc}}$ by increasing the qubit-coupler capacitance will also decrease $E_{\mathrm{Cc}}$ and $E_{\mathrm{Cq}}$ for both qubits. For small systems, we can tune each of these components multiple times with a finite-element method (FEM) simulation to converge towards the target values. However, this approach is relatively slow, requiring $\sim12~\minutes$ per simulation on a desktop computer. Making small changes to the geometry and iterating to get the correct coupling becomes an arduous process, especially for devices with many couplers.

To circumvent this bottleneck, we make use of pre-simulated lookup tables (LUTs) for the transmon and tunable coupler geometries in order to more efficiently calculate the total capacitance matrix. The individual capacitance matrices for each transmons and tunable coupler are calculated first from the LUTs and then combined to form the total capacitance matrix,
\begin{equation*}
\begin{pmatrix}
    \blockmatrix{\qi} & \blockmatrix{\cij\qi}  &   \\
     \blockmatrix{\cij\qi} & \blockmatrix{\cij} & \blockmatrix{\cij\qj}  \\
      &  \blockmatrix{\cij\qj} & \blockmatrix{\qj} \\
\end{pmatrix}.
\end{equation*}

The transmon LUT consists of simulations of the qubit geometry, where most of the geometry is defined by fixed dimensions that are not changed during capacitance targeting [green arrows in Fig.~\ref{fig:lookup_table_tuning}(a)], and free dimensions that are adjusted while targeting a coupling capacitance [golden and red arrows in Fig.~\ref{fig:lookup_table_tuning}(a)]. We use a total of seven adjustable dimensions, six for defining a coupling to a readout resonator or a tunable coupler [golden arrows in Fig.~\ref{fig:lookup_table_tuning}(a)], and one for tuning $E_\mathrm{Cq}$ [red arrow in Fig.~\ref{fig:lookup_table_tuning}(a)]. For each adjustable dimension, two strategic values are chosen for simulation, one at the lower end of possible coupling capacitance and one at the higher end. Importantly, for all the possible dimension values in between, the capacitance has to be highly linear. The linearity ensures that only two values suffice per adjustable dimension for the LUT, making a total of $2^7=128$ required simulations for the full LUT. Comparing the interpolated values with FEM simulations, the error of using the LUT method is estimated to be below $0.3\%$, confirmed with simulations of the charging energy [Fig.~\ref{fig:lookup_table_tuning}(b)]. Crucially, this method decreases tuning times by up to four orders of magnitude.

Connecting the three LUTs introduces a larger error that can reach at most $5\%$ per connection, yielding a maximum error of $4\%$ for our device design [coupler $\ccb$ in Fig.~\ref{fig:lookup_table_tuning}(d)]. Including the second transmon with another connection doubles the prediction error of the LUT method, reaching a maximum value of $8\%$ [coupler $\ccb$ in Fig.~\ref{fig:lookup_table_tuning}(c)]. A possible source of this discrepancy is an unaccounted capacitance from the transmon pad to a connection outside of the transmon square, in this case the part of the tunable coupler pad outside of the transmon pocket. Crucially, the error of the transmon-coupler and transmon-transmon couplings are correlated, making the error in the $\reszz=0$ tunable-coupler frequency less sensitive to LUT prediction error.

\section*{Qubit coherence dependence on coupler bias}

In tunable coupling architectures, the mode used to define the qubit (qubit-like mode) is partially hybridized with the tunable coupler. As the coupler approaches the qubit in frequency, they become more hybridized. Consequently, the coherence of the qubit-like mode decreases as the coupler becomes more sensitive to flux noise [Fig.~\ref{fig:qubit_coherence_coupler}(a,b)].. The Echo decay of the qubit-like modes also becomes more gaussian as the coupler becomes more flux-sensitive [Fig.~\ref{fig:qubit_coherence_coupler}(c)], a signature associated to $1/f$ flux noise~\cite{Braumuller20}. If the qubit and coupler have different relaxation times, the relaxation time of the qubit-like mode will also vary as they become more hybridized [Fig.~\ref{fig:qubit_coherence_coupler}(c)].

\section*{Coupler characterization}

To study the dependence of various parameters on tunable-coupler frequency, it is necessary to be able to drive and readout the tunable couplers. Given that tunable couplers in our device do not have dedicated circuitry for drive and readout, only flux control, we must make use of the neighboring transmons and harness the strong dispersive coupling between coupler and transmons. For example (Fig.~\ref{fig:CouplerSpectroscopy}), pulsed spectroscopy on $\cca$ is implemented by applying a $40~\us$ square pulse via the drive line of $\qa$.

A shorter square pulse of $500~\ns$ is subsequently applied to $\qc$, and $\qc$ then readout~\cite{Marxer23,Zhang23}. This $\qc$ pulse is intended to partially excite the qubit only when the coupler is in the ground state, generating a signal when the $40~\us$ pulse is resonant with $\cca$.
We use this modified pulsed spectroscopy to characterize the flux dependence of $\cca$ frequency. A polynomial fit to this flux arc gives the conversion from static flux to frequency and viceversa. We perform this calibration on all couplers.

The 4 tunable couplers in our device are all strongly coupled to $\qc$. Consequently, there exists a non-negligible interaction between tunable couplers mediated by $\qc$ [Fig.~\ref{fig:CouplerSpectroscopy}(d-e)]. This observation prompts us to characterize each tunable coupler individually, by biasing the non-measured couplers to $\Phi=0.5\Phi_0$ to avoid unwanted population exchange between coupler modes.

The above described spectroscopy method has the advantage of being straightforward to implement, making it convenient in the early device exploration stage. But achieving high contrast between the ground and excited states of the coupler with this method would require longer microwave pulses with a frequency spectrum narrower than the coupler-state-dependent shift in qubit frequency, which in this device is at least $4~\MHz$. As such a longer coupler readout operation could be limited by the coupler relaxation time, we opt for a faster Ramsey-based indirect readout scheme, shown in Fig.~\ref{fig:CouplerReadout}(c). A Ramsey sequence (two $\pihalf$ pulses separated by $\tau$) is applied to one of the neighboring transmons, followed by standard dispersive readout of the transmon. We chose $\tauopt$ to maximize contrast. This method achieves single-shot readout of $\cca$ with $\epsro=8\%$ (Fig.~\ref{fig:CouplerReadout}d), mainly limited by dephasing of $\qc$ during $\tauopt$ and $\qc$ readout fidelity.

This Ramsey-based indirect readout of couplers was made compatible with standard single-qubit calibration routines in Quantify, enabling the characterization of coupler coherence times as displayed in Table~\ref{tab:coupler_summary_table}. It is also compatible with the Cryoscope experiment~\cite{Rol20} used to characterize dynamical distortion of flux pulses applied to the tunable coupler up to $500~\ns$. Furthermore, it is also compatible with the spectroscopy-based flux pulse characterization~\cite{Hellings25} used to measure such distortion up to $30~\us$. By pre-distorting flux pulses using the real-time filtering capability of  QCM modules, we compensate the dynamical distortions in these short and long timescales. This is critical to implementing repeatable, high-fidelity, two-qubit gates with unipolar flux pulses.

\section*{Measurement-induced population exchange in the two-excitation manifold}

The frequency shift and increased linewidth of qubit transitions during readout can induce population exchange between neighboring transmons, as shown in Fig.~\ref{fig:J1ROPExch}. This phenomenon can also occur in the two-excitation manifold. For example, as shown in Fig.~\ref{fig:J2ROPExch}, when $\qc$ and $\qd$ are prepared in $\ket{1_0 1_3}$ and $\qc$ is readout, population exchange to $\ket{0_0 2_3}$ can occur via the same mechanism. This population exchange is avoided when $\ccd$ is biased to $\jtwo^{03}=0$ [Fig.~\ref{fig:J2ROPExch}(b)]. The exchange is small but non-zero at $\reszz^{03}=0$ for the chosen $\qc$ readout-pulse amplitude. We believe this is the reason why in Fig.~\ref{fig:SpectDefRate}(e) the parity-check defect rate with spectator $\qd$ toggling does not perfectly match the defect rate without toggling. Finally, note that as $|\reszz|$ is large when $\jtwo=0$ [Fig.~\ref{fig:J2ROPExch}(c)], we do not consider $\jtwo=0$ a good bias condition for any tunable coupler.

\section{Choosing the readout pulse amplitude for $\qc$}

For the repeated parity-check experiment, the readout on ancilla $\qc$ is realized with a $452~\ns$ square pulse at a frequency optimized to distinguish the ground and first-excited states. The total allocated time for readout is $800~\ns$, long enough to ensure photon depletion in the readout mode~\cite{Bultink16}. Integration of the transmitted signal is performed using optimal weight functions~\cite{Magesan15, Ryan15}. Figure~\ref{fig:DefRateROPulseAmp} shows the dependence of the parity defect rate on $\qc$ readout-pulse amplitude. At low amplitude, the signal-to-noise ratio is worse and the defect rates increase. The defect rate also increases at high readout amplitude and also shows a slope, a signal typically associated to leakage~\cite{Google21} (in this case, measurement induced). Based on this study, we chose the $\qc$ readout-pulse amplitude indicated by the vertical line.

\section*{Fast-adiabatic pulse parametrization}

We implement the baseband flux pulse on the coupler as a unipolar, fast-adiabatic pulse as introduced by Martinis and Geller\cite{Martinis14}, adapted to the tunable coupler system in a similar way as in Ref.~\cite{Sung21b}.
Therefore, we define the time-dependent parameter
\begin{equation}
\theta(t) = \arctan{\left(\frac{\gqc}{\Delta(t)}\right)},
\label{eq:thetadelta}
\end{equation}
\noindent
where $\Delta(t)=\omega_c(t) - \omega_q$ and $\omega_c$ and $\omega_q$ the coupler and high-frequency qubit frequencies, respectively. The starting value, $\theta_i$, is determined by the coupler frequency at the $\jone=0$ bias. The final value, $\theta_f$, is chosen from landscapes similar to those of Figs.~\ref{fig:CZLandscape}(a-b), chosen to minimize leakage along the contour of  $180^\degrees$ conditional phase. Based on these values, the time-dependent pulse is parametrized as

\begin{equation}
\theta(t) = \frac{\theta_f - \theta_i}{2}\left[1-\cos{\left(2\mathrm{\pi}t/t_p\right)}\right],
\end{equation}
\noindent
where $t_p=60~\ns$ is the pulse duration. After generating the $\theta(t)$ pulse, we convert it to coupler frequency using Eq.~\ref{eq:thetadelta} and then into voltage-pulse amplitude using the frequency-to-amplitude conversion characterized using the Cryoscope method~\cite{Rol20} at different pulse amplitudes.

\section*{Defect rate with spectator effects on $\qd$}
The spectator effects observed by toggling $\qe$ during repeated $X_2 X_1$ parity checks (Fig.~\ref{fig:SpectDefRate}) are reproduced when toggling $\qd$ instead. This experiment is shown in Fig.~\ref{fig:SpectDefRateSOM}.

\section*{Emulating residual $ZZ$ errors by flux pulsing the ancilla}

Toggling a spectator like $\qd$ ($\qe$) when $\reszz^{03}\neq0$  $\left(\reszz^{04}\neq0\right)$ forces a frequency shift in $\qc$. To better understand the dynamics of  defect rates under toggling of spectators, we emulate this state-dependent frequency shift of $\qc$ by dynamically detuning $\qc$ an equivalent amount during the two-qubit gates (Fig.~\ref{fig:PulsedDefectRate}). Note that we only apply the dynamical detuning at odd-numbered rounds to emulate the switching of the spectator qubit. We indeed observe that the defect rates also increase with increased dynamical detuning of $\qc$, as expected. However, comparing both experiments at round 100,  we observe that the defect rate increases faster when toggling the spectator than when just dynamically detuning $\qc$ [Fig.~\ref{fig:PulsedDefectRate}(e)]. Moreover, the defect rates measured while toggling a spectator present a more pronounced slope, an effect typically associated with leakage to higher excited transmon states~\cite{Google21}. We surmise that the root of the discrepancy may be that toggling the spectator affects the hybridization of $\qc$, not just its frequency.

\bibliographystyle{apsrev4-2}

\end{bibunit}

\end{document}